\documentclass[12pt]{article}
\usepackage{epsfig}
\usepackage{color}
\usepackage{amssymb,amsmath}

\newcommand{\bea}{\begin{eqnarray}}
\newcommand{\eea}{\end{eqnarray}}
\newcommand{\be}{\begin{equation}}
\newcommand{\ee}{\end{equation}}

 \makeatletter
\def\alt{\mathrel{\mathpalette\gl@align<}}
\def\agt{\mathrel{\mathpalette\gl@align>}}
\def\gl@align#1#2{\lower.6ex\vbox{\baselineskip\z@skip\lineskip\z@
\ialign{$\m@th#1\hfil##\hfil$\crcr#2\crcr\sim\crcr}}} \makeatother
\begin{document}
\begin{flushright}
\end{flushright}
\vspace*{1.0cm}

\begin{center}
\baselineskip 20pt 
{\Large\bf 
R-parity Conserving Minimal SUSY $B-L$ Model 
}
\vspace{1cm}

{\large 
Nobuchika Okada and Nathan Papapietro
}
\vspace{.5cm}

{\baselineskip 20pt \it
Department of Physics and Astronomy, 
University of Alabama, 
Tuscaloosa,  AL 35487, USA
} 

\vspace{.5cm}

\vspace{1.5cm} {\bf Abstract}\\
\end{center}

We propose a simple gauged U(1)$_{B-L}$ extension of the minimal supersymmetric Standard Model (MSSM), 
   where R-parity is conserved as usual in the MSSM. 
The global $B-L$ (baryon number minus lepton number) symmetry in the MSSM is gauged and three MSSM gauge-singlet 
   chiral multiplets with a unit $B-L$ charge are introduced, 
   ensuring the model free from gauge and gravitational anomalies. 
We assign an odd R-parity for two of the new chiral multiplets 
  and hence they are identified with the right-handed neutrino superfields, 
  while an even R-parity is assigned to the other one ($\Phi$). 
The scalar component of $\Phi$  plays the role of a Higgs field 
  to break the U(1)$_{B-L}$ symmetry through its negative mass squared, 
  which is radiatively generated by the renormalization group running of soft 
  supersymmetry (SUSY) breaking parameters from a high energy. 
This radiative U(1)$_{B-L}$ symmetry breaking leads to its breaking scale being at the TeV naturally.  
Because of our novel R-parity assignment, three light neutrinos are Dirac particles with one massless state. 
Since R-parity is conserved, the lightest neutralino is a prime candidate of the dark matter as usual. 
In our model, the lightest eigenstate of the mixture of the $B-L$ gaugino and 
  the fermionic component of $\Phi$ appears as a new dark matter candidate. 
We investigate phenomenology of this dark matter particle. 
We also discuss collider phenomenology of our model. 
In particular, the $B-L$ gauge boson ($Z'$), once discovered at the Large Hadron Collider, 
  can be a probe to determine the number of (right-handed) Dirac neutrinos with its invisible decay width,  
  in sharp contrast with the conventional $B-L$ extension of the SM or MSSM  
  where the right-handed neutrinos are heavy Majorana particles and decay to the SM leptons. 

\thispagestyle{empty}

\newpage

\addtocounter{page}{-1}
\setcounter{footnote}{0}
\baselineskip 18pt
\section{Introduction}

The $B-L$ (baryon number minus lepton number)  
  is the unique anomaly-free global U(1)$_{B-L}$ symmetry in the Standard Model (SM). 
This symmetry is easily gauged, and the so-called minimal $B-L$ model~\cite{MBL1}-\cite{MBL6} 
  is a simple gauged $B-L$ extension of the SM, where three right-handed neutrinos 
  and an SM gauge singlet Higgs field with two units of the $B-L$ charge are introduced. 
The three right-handed neutrinos are necessarily introduced to make the model 
  free from all gauge and gravitational anomalies.  
Associated with a $B-L$ symmetry breaking by a Vacuum Expectation Value (VEV) 
  of the $B-L$ Higgs field, the $B-L$ gauge field ($Z^\prime$ boson) and 
  the right-handed neutrinos acquire their masses. 
After the electroweak symmetry breaking, tiny SM neutrino masses are generated 
  via the seesaw mechanism~\cite{seesaw1}-\cite{seesaw5}.

Although the scale of the $B-L$ gauge symmetry breaking is arbitrary 
  as long as phenomenological constraints are satisfied, 
  a breaking at the TeV scale is probably the most 
  interesting possibility in the view point of the Large Hadron Collider (LHC) experiments. 
However, mass squared corrections of the $B-L$ Higgs (any Higgs fields in 4-dimensional models, in general) 
  are quadratically sensitive to the scale of a possible ultraviolet theory, and as a result 
  the $B-L$ symmetry breaking scale is unstable against quantum corrections.   
As is well-known, supersymmetric (SUSY) extension is the most promising way 
  to solve this vacuum instability.  
Very interestingly, SUSY extension of the minimal $B-L$ model offers a way 
  to naturally realize the $B-L$ symmetry breaking at the TeV scale.  
With suitable inputs of soft SUSY breaking parameters at a high energy, 
  their renormalization group (RG) evolutions drive the $B-L$ Higgs mass squared 
  negative and therefore the $B-L$ gauge symmetry is radiatively broken~\cite{Khalil:2007dr, FileviezPerez:2010ek, Burell:2011wh}.  
Since the scale of the negative mass squared is controlled by the soft SUSY breaking parameters, 
  the $B-L$ breaking scale lies at the TeV from naturalness.

SUSY extension opens a further possibility. 
As has been proposed in Ref.~\cite{Barger:2008wn}, 
  it is not necessary to introduce the $B-L$ Higgs field, 
  since the scalar partner of a right-handed neutrino can play the same role 
  as the $B-L$ Higgs field in breaking the $B-L$ gauge symmetry.  
Hence, we can define the ``minimal SUSY $B-L$ model'' by a particle content, 
  where only three right-handed neutrino chiral superfields are added 
  to the particle content of the minimal SUSY SM (MSSM). 
It is interesting that such a particle content can be derived from heterotic strings~\cite{Braun:2005nv, Anderson:2009mh}. 
In Ref.~\cite{Barger:2008wn}, a negative soft mass squared of a right-handed sneutrino 
  is assumed to break the $B-L$ gauge symmetry, so that the $B-L$ symmetry 
  breaking occurs at the TeV scale.  
Associated with this symmetry breaking, R-parity is also spontaneously broken, 
  and many interesting phenomenologies with the R-parity violation 
  have been discussed~\cite{FileviezPerez:2009gr,  Everett:2009vy, FileviezPerez:2012mj, Marshall:2014kea}.  
Through the non-zero VEV of  the right-handed sneutrino, mixings between neutrinos, 
  MSSM Higgsinos, MSSM neutralinos and $B-L$ gaugino are generated. 
Although the neutrino mass matrix becomes very complicated, 
  it has enough number of degrees of freedom to reproduce the neutrino oscillation data 
  with a characteristic pattern of the mass spectrum~\cite{Barger:2010iv, Ghosh:2010hy}.

In this paper,  we propose the minimal SUSY $B-L$ model with an R-parity conservation.  
The particle content is the same as the one of the minimal SUSY $B-L$ model discussed above, 
  while we assign an even R-parity to one right-handed neutrino chiral superfield ($\Phi$)
  and an odd  R-parity to the other two right-handed neutrino chiral superfields. 
The R-parity assignment for the MSSM fields is as usual. 
Because of this parity assignment and the gauge symmetry, the chiral superfield $\Phi$ 
  has no Dirac Yukawa coupling with the lepton doublet fields.  
In fact, it does not appear in the renormalizable superpotential.  
We consider the case that the $B-L$ symmetry breaking is driven by a VEV of the R-parity even right-handed sneutrino. 
Phenomenological consequences in this model are very different from those of the conventional minimal SUSY $B-L$ model. 
As usual in the MSSM, R-parity is conserved and hence the lightest neutralino is a candidate of the dark matter. 
In addition to the lightest neutralino in the MSSM, the model offers a new candidate for the dark matter, 
  namely, a linear combination of  the fermion component of $\Phi$ and the $B-L$ gaugino.   
Since $\Phi$ has no Dirac Yukawa coupling, no Majorana  mass term is generated in the SM neutrino sector, 
  and as a result, the SM neutrinos are Dirac particles. 
With only the two right-handed neutrinos involved in the Dirac Yukawa couplings, 
  the Dirac neutrino mass matrix leads to three mass eigenstates, 
  one massless chiral neutrino and two Dirac neutrinos. 
A general 2-by-3 Dirac mass matrix includes a number of free parameters enough to reproduce the neutrino oscillation data. 
This Dirac nature of the SM neutrinos are quite distinctive from those in the usual $B-L$ model, 
  where the right-handed neutrinos are heavy Majorana particles and the mass eigenstates different 
  from the light SM neutrinos. 
If the $Z^\prime$ boson is discovered at the LHC, this difference could be tested  
  through its decay products and the decay width measurements.

This paper is organized as follows. 
In the next section, we define our minimal SUSY $B-L$ model 
  with a novel R-parity assignment. 
Then, we introduce the superpotential and soft SUSY breaking terms 
  relevant for our discussion. 
In Sec.~3, we discuss a way to radiatively break the $B-L$ gauge symmetry,  
   while keeping R-parity manifest.  
Focusing on the $B-L$ sector, for simplicity, we perform a numerical analysis 
  for the RG evolutions of the soft SUSY breaking masses 
  of the right-handed sneutrinos, and show that the $B-L$ gauge symmetry 
  is radiatively broken at the TeV scale by a VEV of the scalar component of $\Phi$. 
In Sec.~4, we consider a new dark matter candidate which is a linear combination of 
   the scalar component of $\Phi$ and the $B-L$ gaugino.  
We show a parameter set which can reproduce the observed dark matter relic density. 
In Sec.~5, we also briefly discuss an implication of the Dirac neutrinos to the LHC phenomenology 
   through the $Z^\prime$ boson production. 
The last section is devoted for conclusions and discussions.

\section{Minimal SUSY $B-L$ model with a conserved R-parity}

\begin{table}[t]
\begin{center}
\begin{tabular}{|c||ccc|c|c|c|}
\hline
chiral superfield & SU(3)$_c$ & SU(2)$_L$ & U(1)$_Y$ 
 & U(1)$_{B-L}$ & R-parity  \\
\hline \hline
$ Q^i   $  & {\bf 3}     & {\bf 2} & $+1/6$ & $+1/3$ & $-$    \\ 
$ U^c_i $  & {\bf 3}$^*$ & {\bf 1} & $-2/3$ & $-1/3$ & $-$  \\ 
$ D^c_i $  & {\bf 3}$^*$ & {\bf 1} & $+1/3$ & $-1/3$ & $-$ \\ 
\hline
$ L_i   $    & {\bf 1} & {\bf 2}& $-1/2$ & $-1$   & $-$   \\ 
$ \Phi   $  & {\bf 1} & {\bf 1}& $  0 $ & $+1$   & $+$   \\ 
$ N^c_{1,2} $  & {\bf 1} & {\bf 1}& $  0 $ & $+1$ & $-$  \\ 
$ E^c_i $  & {\bf 1} & {\bf 1}& $ -1 $ & $+1$     & $-$  \\ 
\hline 
$ H_u$     & {\bf 1} & {\bf 2}   & $+1/2$ &  $ 0$ & $+$  \\ 
$ H_d$     & {\bf 1} & {\bf 2}   & $-1/2$ &  $ 0$ & $+$   \\  
\hline
\end{tabular}
\end{center}
\caption{
Particle content of the minimal SUSY $B-L$ model with a conserved R-parity.    
In addition to the MSSM particles, 
  three right-handed neutrino superfields ($\Phi$ and $N^c_{1,2}$) are introduced. 
We assign an even R-parity for $\Phi$. 
$i=1,2,3$ is the generation index. 
}
\label{BLcontent}
\end{table}

The minimal SUSY $B-L$ model is based on the gauge group of  
  SU(3)$_c\times$SU(2)$_L \times $U(1)$_Y \times$U(1)$_{B-L}$.  
In addition to the MSSM particle content, we introduce 
  three chiral superfields which are singlets under the SM gauge groups  
  and have a unit $B-L$ charge. 
The new fields are identified as the right-handed neutrino chiral superfields, 
  and their existence is essential to make the model free from all gauge and gravitational anomalies.   
Unlikely to direct supersymmetrization of the minimal $B-L$ model, 
  the $B-L$ Higgs superfields are not included in the particle content. 
The key of our proposal is that we assign an even R-parity 
  to one right-handed neutrino chiral superfield, 
  in contrast with the minimal SUSY $B-L$ model proposed in Ref.~\cite{Barger:2008wn},
  where all the right-handed neutrino superfields are R-parity odd as usual. 
The particle content is listed in Table~\ref{BLcontent}.

The gauge and parity invariant superpotential which is added to the MSSM one is 
   only the neutrino Dirac Yukawa coupling as 
\begin{eqnarray}
 W_{BL} = \sum_{i=1}^2 \sum_{j=1}^3 y_D^{ij} N^c_i L_j H_u.  
\label{WBL} 
\end{eqnarray}
Note that the Yukawa coupling for $\Phi$ is forbidden by the parity, 
   and $\Phi$ has no direct coupling with the MSSM fields.  
After the electroweak symmetry breaking, the neutrino Dirac mass matrix is generated. 
Since this is a 2-by-3 matrix, one neutrino remains massless.  
Therefore, we have one massless neutrino and two Dirac neutrinos in the model. 
The 2-by-3 Dirac mass matrix has a sufficient number of free parameters 
  to reproduce the neutrino oscillation data. 
Although we have introduced the special parity assignment, 
  this may be unnecessary in the practical point of view. 
Without the parity assignment, the superpotential in Eq.~(\ref{WBL}) can include 
\begin{eqnarray}
   W_{BL}  \supset \sum_{j=1}^3 y_D^j \Phi L_j H_u, 
\end{eqnarray}
which are unique direct couplings between $\Phi$ and the MSSM fields. 
Let us now take a limit $y_D^j \to 0$, which switch off the direct communication of $\Phi$  
   with the lepton and Higgs doublets.   
In this sense, our parity assignment can be regarded as a result of symmetry enhancement 
   caused by this limit. 
Since the neutrinos are Dirac particles, the Dirac Yukawa coupling constants must be extremely small 
  in order to reproduce the observed neutrino mass scale. 
We will discuss a possibility to naturally realize such small parameters in the last section.

Next, we introduce soft SUSY breaking terms for the fields in the $B-L$ sector: 
\begin{eqnarray}
  {\cal L}_{\rm soft}&=& 
- \left( \frac{1}{2} M_{BL} \lambda_{BL} \lambda_{BL} + h.c.  \right)
- \left( \sum_{i=1}^2 m_{\tilde{N}^c_i}^2 |\tilde{N^c_i}|^2 
+ m_\phi^2 |\phi|^2   \right) , 
\end{eqnarray}
where $\lambda_{BL}$ is the $B-L$ gaugino and 
  $\tilde{N^c_i}$ and $\phi$ are scalar components of $N^c_i$ and $\Phi$, respectively. 
Since the Dirac Yukawa couplings are very small, we omit terms relevant to the couplings. 
In the next section, we analyze the RG evolutions of the soft SUSY breaking masses 
  and find that $m_\phi^2$ is driven to be negative and the U(1)$_{B-L}$ symmetry is radiatively broken. 
Although we do not assume the grand unification of our model, 
  we take $M_U =2 \times 10^{16}$ GeV as a reference scale 
  at which the boundary conditions for the soft masses are given.

\section{Radiative $B-L$ symmetry breaking} 

\begin{figure}[t]
\begin{center}
{\includegraphics[scale=1.5]{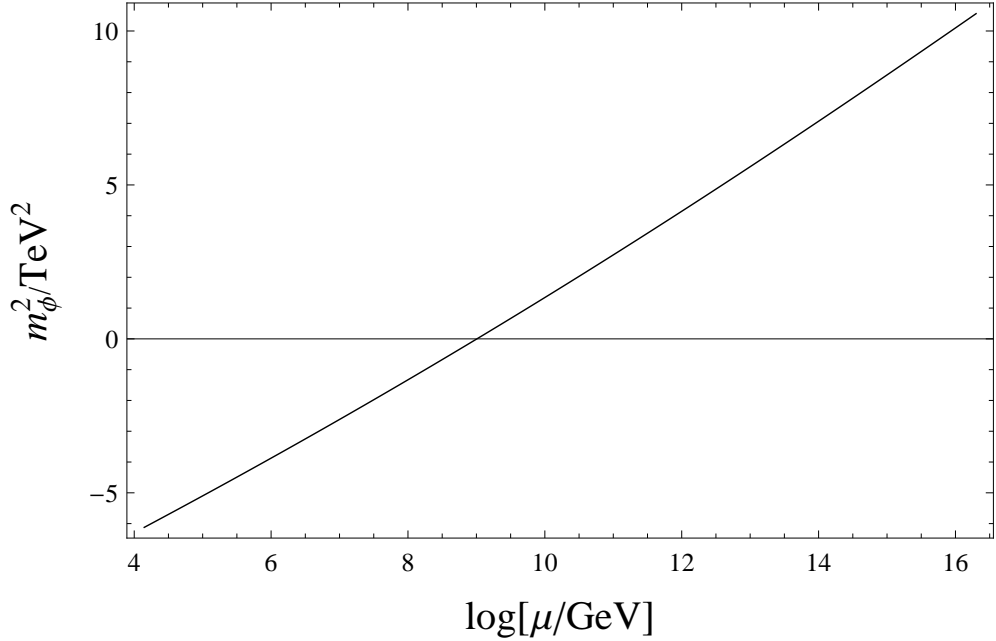}}
\caption{
The RG evolution of the soft SUSY breaking mass $m_\phi^2$ from $M_U$ to low energies. 
}
\label{fig:mphiRun}
\end{center}
\end{figure}

It is well-known that the electroweak symmetry breaking in the MSSM 
   is triggered by radiative corrections which drive soft mass squared 
   of the up-type Higgs doublet negative. 
Because of this radiative symmetry breaking, the electroweak symmetry breaking scale 
   is controlled by the soft SUSY breaking mass scale and the SUSY breaking scale at the TeV 
   naturally results in the right electroweak scale of ${\cal O}$(100 GeV). 
Similarly to the MSSM, a radiative $B-L$ symmetry breaking occurs by the RG evolution  
  of soft SUSY breaking parameters  from a high energy to low energies.  
However, the mechanism that drives $m_\phi^2$ negative is different 
  from the one in the MSSM where the large top Yukawa coupling plays a crucial role.

To make our discussion simple, we consider the RG equations only for the $B-L$ sector.\footnote{
See Refs.~\cite{Ambroso:2010pe, Ovrut:2015uea}  for more elaborate analysis and parameter scans 
  to identify parameter regions which are consistent with current experimental results.}
RG equations relevant for our discussion are 
\begin{eqnarray} 
\label{BL-RGE1}
 16 \pi^2 \mu \frac{d M_{BL}}{d \mu} 
&=&  32 g_{BL}^2 M_{BL}, \\ 
\label{BL-RGE2}
 16 \pi^2 \mu \frac{d m_{\tilde{N}^c_i}^2}{d \mu} 
&=&  - 8 g_{BL}^2 M_{BL}^2 + 2 g_{BL}^2 \left(  \sum_{j=1}^2 m_{\tilde{N}^c_j}^2 + m_\phi^2  \right),  \\
\label{BL-RGE3}
 16 \pi^2 \mu \frac{d m_\phi^2}{d \mu} 
&=&  - 8 g_{BL}^2 M_{BL}^2 + 2 g_{BL}^2 \left(  \sum_{j=1}^2 m_{\tilde{N}^c_j}^2 + m_\phi^2  \right),  
\end{eqnarray} 
 where the $B-L$ gauge coupling obeys 
\begin{eqnarray}
  16 \pi^2 \mu \frac{d g_{BL}}{d \mu}  =  16 g_{BL}^3. 
\end{eqnarray} 
In Eq.~(\ref{BL-RGE2}) the contributions from very small Dirac Yukawa couplings are omitted. 
In fact, the second term in the right-hand side of Eq.~(\ref{BL-RGE3}), 
   which originates from the $D$-term interaction, 
   plays an essential role to drive $m_\phi^2$  negative.  
Since squarks and sleptons have $B-L$ charges, their soft squared masses also appear 
   in the RG equations, but we have omitted them, for simplicity, by assuming they are 
   much smaller than $m_{\tilde{N}^c_i}^2$ and $m_\phi^2$.

To illustrate the radiative $B-L$ symmetry breaking, 
  we numerically solve the above RG equations from $M_U=2 \times 10^{16}$ GeV 
  to low energy, choosing the following boundary conditions. 
\begin{eqnarray}
 g_{BL}= 0.311, \; \;  M_{BL}=8.13 \; {\rm TeV}, \; \; 
   m_{\tilde{N}^c_1} = m_{\tilde{N}^c_2} = 20.0 \; {\rm TeV}, \; \; 
   m_{\phi} = 3.25 \; {\rm TeV}.  
\label{BC}
\end{eqnarray}
Fig.~\ref{fig:mphiRun} shows the RG evolution of $m_\phi^2$. 
The mass squared of $\phi$ becomes negative at low energies as shown in this figure, 
   while the other squared masses remain positive. 
The mass squared hierarchy $m_{\tilde{N}^c_i} \gg m_\phi^2$ is crucial to drive $m_\phi^2 < 0$.

We now analyze the scalar potential with the soft SUSY breaking parameters obtained 
   from the RG evolutions. 
We choose the VEV of $\phi$ as $v_{BL}=\sqrt{2} \langle \phi \rangle = 14$ TeV as a reference, 
   at which the solutions of the RG equations are evaluated as follows: 
\begin{eqnarray}
 g_{BL}= 0.250, \; \;  M_{BL}=5.25 \; {\rm TeV}, \; \; 
   m_{\tilde{N}^c_1} = m_{\tilde{N}^c_2} = 19.6 \; {\rm TeV}, \; \; 
   |m_{\phi}| = 2.47 \; {\rm TeV}.  
\label{LEmass}
\end{eqnarray}
The scalar potential is given by 
\begin{eqnarray}
  V = m_{\tilde{N}^c_1}^2 |\tilde{N}^c_1|^2   +  m_{\tilde{N}^c_2}^2 |\tilde{N}^c_2|^2 + m_{\phi}^2 |\phi|^2
       + \frac{g_{BL}^2}{2} 
       \left(  |\tilde{N}^c_1|^2  + |\tilde{N}^c_2|^2  + |\phi|^2  \right)^2. 
\end{eqnarray}
Solving the stationary conditions, we find (in units of TeV)  
\begin{eqnarray}
 \langle \tilde{N}^c_1 \rangle = \langle \tilde{N}^c_2 \rangle =0, \; \; 
 \langle \phi \rangle =\frac{\sqrt{-2 m_\phi^2}}{g_{BL}} \simeq \frac{14}{\sqrt 2}  . 
\end{eqnarray} 
This result is consistent with our choice of $v_{BL}=14$ TeV in evaluating the running soft masses.

In our parameter choice, the $Z^\prime$ boson mass is given by 
\begin{eqnarray}  
   m_{Z^\prime} = g_{BL} v_{BL} = 3.5 \; {\rm TeV}. 
\end{eqnarray}  
The ATLAS and CMS collaborations at the LHC Run-2 have been searching 
  for the $Z^\prime$ boson resonance with the dilepton final state 
  and have recently reported their results which are consistent with the SM expectaions~\cite{ATLAS-Zp, CMS:2015nhc}.   
In Ref.~\cite{Okada:2016gsh}, the ATLAS and CMS search results are interpreted to 
   a constraint on the $Z^\prime$ boson in the minimal $B-L$ model, 
   where an upper bound of the the $B-L$ gauge coupling 
   as a function of  $Z^\prime$ boson mass has been obtained. 
We refer the results in Ref.~\cite{Okada:2016gsh} such that 
   $g_{BL} \leq 0.328$ and $0.350$ for $m_{Z^\prime}=3.5$ TeV from the ATLAS and CMS results, respectively.\footnote{
It is also shown in Ref.~\cite{Okada:2016gsh} that the ATLAS bound at the LHC Run-2 
   is more severe than the bound obtained from the LEP2 data~\cite{Schael:2013ita} for $m_{Z^\prime} \leq 4.3$ TeV.  
}   
Our parameter choice of $g_{BL}=0.250$ for $m_{Z^\prime}=3.5$ TeV 
   is consistent with the recent LHC Run-2 results.

\section{Right-handed neutrino dark matter} 
As we showed in the previous section, the $B-L$ gauge symmetry is radiatively broken 
   by the RG effects on the soft SUSY breaking masses. 
Since the breaking occurs by the VEV of R-parity even scalar field $\phi$, 
   R-parity is still manifest, by which the stability of the lightest R-parity odd particle is ensured. 
Thus, as usual in the MSSM, the lightest neutralino is a candidate of the dark matter. 
In addition to the MSSM neutralinos, a new dark matter candidate arises in our model, 
   namely, the fermion component of $\Phi$ ($\psi$).  
We can call $\psi$ the R-parity odd right-handed neutrino. 
In this section, we study phenomenology of the right-handed neutrino dark matter.

A scenario of the parity-odd right-handed Majorana neutrino dark matter was first proposed in \cite{Okada:2010wd} 
   in the context of the non-SUSY minimal $B-L$ model, 
   where a $Z_2$ parity is introduced and an odd parity is assigned to one right-handed neutrino 
   while the other fields are all parity-even. 
Because of the $Z_2$-parity conservation, the parity-odd right-handed neutrino becomes stable 
   and hence the dark matter candidate.   
Phenomenology of this dark matter has been investigated in \cite{Okada:2010wd, Okada:2012sg, Basak:2013cga}.  
Recently, in terms of the complementarity to the LHC physics, 
   the right-handed neutrino dark matter has been investigated in detail~\cite{Okada:2016gsh}. 
Supersymmetric version of the minimal $B-L$ model with the right-handed neutrino dark matter 
  has been proposed in \cite{Burell:2011wh}.

Our dark matter scenario that we will investigate in this section shares similar properties 
   with the scenario discussed in \cite{Burell:2011wh}.  
However, there is a crucial difference that $\psi$ has no Majorana mass by its own, 
   but it acquires a Majorana mass through a mixing with the $B-L$ gaugino ($\lambda_{BL}$). 
After the U(1)$_{B-L}$ symmetry breaking, a mass matrix for $\psi$ and $\lambda_{BL}$ is generated to be 
\begin{eqnarray}
M_\chi=\left(
\begin{array}{cc}
    0                      &   m_{Z^\prime}   \\
    m_{Z^\prime}   &   M_{BL}     
\end{array}
\right). 
\end{eqnarray} 
The mass matrix is diagonalized as 
\begin{eqnarray}
\left(
\begin{array}{c}
    \psi \\
    \lambda_{BL}
\end{array}
\right) =
\left(
\begin{array}{cc}
    \cos \theta       &   \sin \theta  \\
    -\sin \theta       &   \cos \theta       
\end{array}
\right)
\left(
\begin{array}{c}
    \chi_\ell \\
    \chi_h    
\end{array}
\right)
\end{eqnarray} 
with $\tan 2 \theta= 2 m_{Z^\prime}/M_{BL}$.  
Let us assume that the lighter mass eigenstate ($\chi_\ell$) is the lightest neutralino. 
Since $\psi$ and $\lambda_{BL}$ are the SM gauge singlets, possible annihilation processes 
   of the dark matter are very limited. 
Furthermore, given a small $B-L$ gauge coupling and the Majorana nature of the dark matter particle, 
   the annihilation process via sfermion exchanges is not efficient.  
We find that a pair of dark matter particles can annihilate efficiently 
   only if the dark matter mass is close to half of the $Z^\prime$ boson mass 
   and the $Z^\prime$ boson resonance in the $s$-channel annihilation process 
   enhances the cross section.   
Let us set $M_{BL} \simeq (3/2) m_{Z^\prime}$, so that the lightest mass eigenvalue is 
   found to be $m_{DM} \simeq m_{Z^\prime}/2$ and $\cos^2 \theta \simeq 0.8$. 
Our parameter choice in the previous section is suitable for this setup, 
  $M_{BL}= (3/2) m_{Z^\prime}=5.25$ TeV for $m_{Z^\prime}=3.5$ TeV.

Let us now calculate the dark matter relic abundance by integrating the Boltzmann equation given by 
\bea 
  \frac{dY}{dx}
  = - \frac{s \langle \sigma v \rangle}{x H(m_{DM})} \left( Y^2-Y_{EQ}^2 \right), 
\label{Boltmann}
\eea  
where temperature of the universe is normalized by the mass of the right-handed neutrino $x=m_{DM}/T$, 
   $H(m_{DM})$ is the Hubble parameter at $T=m_{DM}$, 
   $Y$ is the yield (the ratio of the dark matter number density to the entropy density $s$) of 
  the dark matter particle, $Y_{EQ}$ is the yield of the dark matter particle in thermal equilibrium, 
  and $\langle \sigma v \rangle$ is the thermal average of the dark matter annihilation cross section times relative velocity. 
Explicit formulas of the quantities involved in the Boltzmann equation are as follows: 
\bea 
s &=& \frac{2  \pi^2}{45} g_\star \frac{m_{DM}^3}{x^3} ,  \nonumber \\
H(m_{DM}) &=&  \sqrt{\frac{4 \pi^3}{45} g_\star} \frac{m_{DM}^2}{M_{Pl}},  \nonumber \\ 
s Y_{EQ}&=& \frac{g_{DM}}{2 \pi^2} \frac{m_{DM}^3}{x} K_2(x),   
\eea
where $M_{Pl}=1.22 \times 10^{19}$  GeV is the Planck mass, 
   $g_{DM}=2$ is the number of degrees of freedom for the Majorana dark matter particle, 
   $g_\star$ is the effective total number of degrees of freedom for particles in thermal equilibrium 
   (in the following analysis, we use $g_\star=106.75$ for the SM particles),  
   and $K_2$ is the modified Bessel function of the second kind.   
In our scenario, a pair of dark matter annihilates into the SM particles 
   dominantly through the $Z^\prime$ boson exchange in the $s$-channel.  
The thermal average of the annihilation cross section is given by 
\bea 
\langle \sigma v \rangle = \left(s Y_{EQ} \right)^{-2} 
  \frac{m_{DM}}{64 \pi^4 x} 
  \int_{4 m_{DM}^2}^\infty  ds \; \hat{\sigma}(s) \sqrt{s} K_1 \left(\frac{x \sqrt{s}}{m_{DM}}\right) , 
\label{ThAvgSigma}
\eea
where the reduced cross section is defined as $\hat{\sigma}(s)=2 (s- 4 m_{DM}^2) \sigma(s)$ 
   with the total annihilation cross section $\sigma(s)$, and $K_1$ is the modified Bessel function of the first kind. 
The total cross section of the dark matter annihilation process $\chi_\ell \chi_\ell \to Z^\prime \to f {\bar f}$ 
  ($f$ denotes the SM fermions plus two right-handed neutrinos) 
   is calculated as 
\bea 
 \sigma(s)= \frac{5}{4 \pi} g_{BL}^4  \cos^2 \theta  \frac{\sqrt{s (s-4 m_{DM}^2)}}
  {(s-m_{Z^\prime}^2)^2+m_{Z^\prime}^2 \Gamma_{Z^\prime}^2} ,
\eea 
where all final state fermion masses have been neglected. 
The total decay width of $Z^\prime$ boson is given by 
\bea
\Gamma_{Z'} = 
 \frac{g_{BL}^2}{24 \pi} m_{Z^\prime} 
 \left[ 15 + \cos^2 \theta  \left( 1-\frac{4 m_{DM}^2}{m_{Z^\prime}^2} \right)^{\frac{3}{2}} 
 \theta \left( \frac{m_{Z^\prime}^2}{m_{DM}^2} - 4 \right)  \right]. 
\label{width}
\eea
Here, we have assumed that all sparticles have mass larger than $m_{Z^\prime}/2$. 
 
\begin{figure}[t]
\begin{center}
{\includegraphics[scale=1.3]{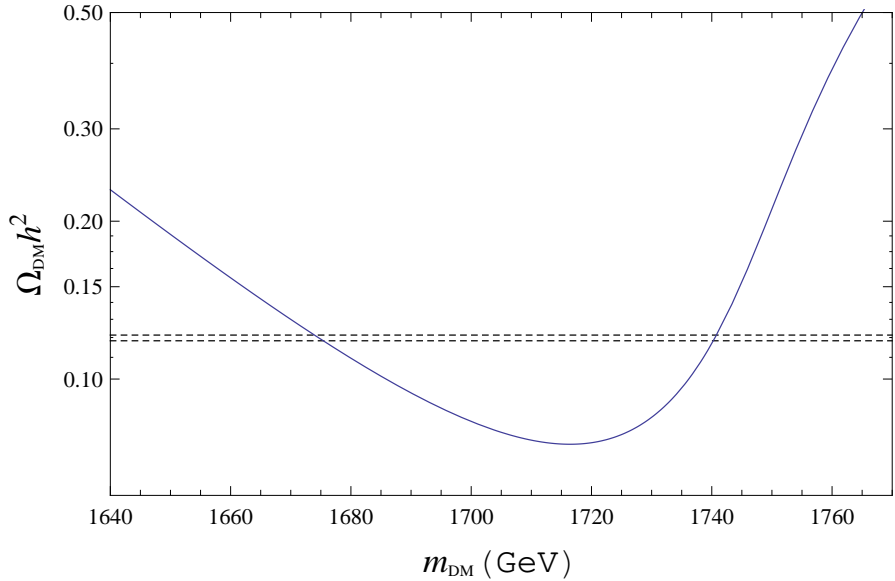}}
\caption{
The relic abundance of the dark matter particle  
  as a function of the dark matter mass ($m_{DM}$) for 
  $g_{BL}=0.250$, $m_{Z^\prime}=3.5$ TeV and $\cos^2 \theta=0.8$. 
The two horizontal lines denote the range of the observed dark matter relic density, 
  $0.1183 \leq \Omega_{DM} h^2 \leq 0.1213$~\cite{Aghanim:2015xee}. 
}
\label{fig:relic}
\end{center}
\end{figure}

Now we solve the Boltzmann equation numerically, and find the asymptotic value of the yield $Y(\infty)$. 
The dark matter relic density is evaluated as 
\bea 
  \Omega_{DM} h^2 =\frac{m_{DM} s_0 Y(\infty)} {\rho_c/h^2}, 
\eea 
  where $s_0 = 2890$ cm$^{-3}$ is the entropy density of the present universe, 
  and $\rho_c/h^2 =1.05 \times 10^{-5}$ GeV/cm$^3$ is the critical density.
In our analysis, only three parameters, 
   namely $g_{BL}$, $m_{Z^\prime}$ and $m_{DM}$, are involved.\footnote{
The mixing angle $\theta$  is determined once $m_{Z^\prime}$ and $m_{DM}$ are fixed.
}
As mentioned above, a sufficiently large annihilation cross section is achieved 
   only if $m_{DM} \simeq m_{Z^\prime}/2$. 
Thus, we focus on the dark matter mass in this region and in this case $\cos^2 \theta \simeq 0.8$. 
For $g_{BL}=0.250$, $m_{Z^\prime}=3.5$ TeV and $\cos^2 \theta=0.8$, 
  Fig.~\ref{fig:relic} shows the resultant dark matter relic abundance 
  as a function of the dark matter mass $m_{DM}$, 
  along with the bound $0.1183 \leq \Omega_{DM} h^2 \leq 0.1213$ (65\% limit) 
  from the Planck satellite experiment~\cite{Aghanim:2015xee} 
  (two horizontal dashed lines). 
We have confirmed that only if the dark matter mass is close to half of the $Z^\prime$ boson mass, 
  the observed relic abundance can be reproduced.

\section{Implication of Dirac neutrino to LHC physics}

Because of our R-parity assignment, the SM neutrinos are Dirac particles in our model.  
This is quite distinct from usual $B-L$ extension of the SM, where right-handed neutrinos are heavy Majorana states. 
Since the right-handed neutrinos are singlet under the SM gauge groups and 
   the Dirac Yukawa coupling constants are very small in both Dirac and Majorana cases, 
   the right-handed neutrinos can communicate with the SM particles only through $Z^\prime$ boson exchange.   
As we mentioned above, the search for $Z^\prime$ boson resonance is underway at the LHC Run-2. 
Once discovered at the LHC, the $Z^\prime$ boson will allows us to investigate 
   physics of the right-handed neutrinos through precise measurements of $Z^\prime$ boson properties. 
In this section we consider an implication of the Dirac neutrinos to LHC physics.

When the right-handed neutrinos are heavy Majorana particles as in the minimal $B-L$ model, 
    a pair of right-handed Majorana neutrinos, if kinematically allowed, can be produced 
    through $Z^\prime$ boson decays at the LHC. 
The produced right-handed neutrino subsequently decays to weak gauge bosons/Higgs boson plus leptons. 
Because of the Majorana nature of the right-handed neutrino, the final states include same-sign leptons. 
This is a characteristic signature from the lepton number violation, and we expect a high possibility 
    to detect such final states with less SM background.  
For a detailed studies, see, for example, Ref.~\cite{Huitu:2008gf}.

The Majorana neutrinos are heavy and can be produced only if they are kinematically allowed, 
     while the Dirac neutrinos in our model are always included in the $Z^\prime$ boson decay products. 
However, they cannot be detected just like the usual SM neutrinos produced at colliders. 
This process may remind us of the neutrino production at the LEP through the resonant production of the $Z$ boson. 
It was a great success of the LEP experiment that the precise measurement of the $Z$ boson decay width 
    and the production cross section at energies around the $Z$ boson peak  
    has determined the number of the SM neutrinos to be three~\cite{ALEPH:2005ab}. 
We notice that the $Z^\prime$ production is quite analogous to the $Z$ production at the LEP. 
Although the right-handed neutrinos produced by the $Z^\prime$ boson are completely undetectable, 
    the total $Z^\prime$ boson decay width carries the information of the invisible decay width.  
A precise measurement of the $Z^\prime$ boson cross section at the LHC may reveal 
    the existence of the right-handed Dirac neutrinos. 
To illustrate this idea, we calculate in the following the differential cross section for the process 
    with the dilepton final states,  
    $pp \to \ell^+ \ell^-$ with $\ell=e, \mu$ mediated by photon, $Z$ boson and $Z^\prime$ boson 
    at the LHC with a collider energy $\sqrt{s}=14$ TeV.

\begin{figure}[t]
\begin{center}
\includegraphics[scale=1.6]{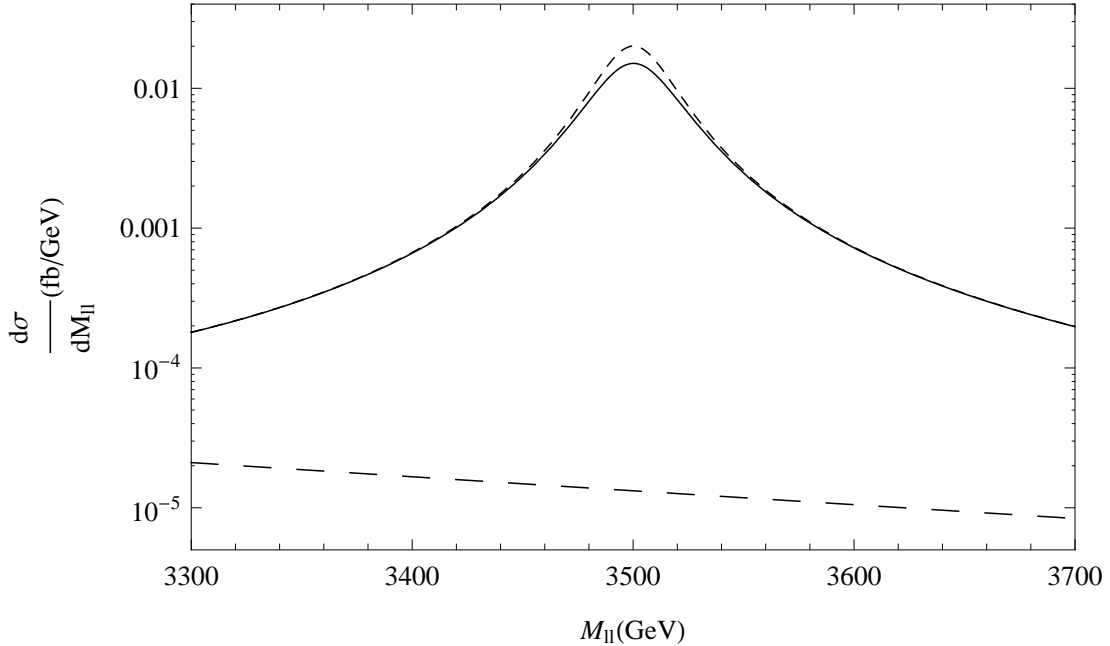}
\caption{
The differential cross section for $pp \to e^+ e^- X  + \mu^+ \mu^- X $ 
  at the 14 TeV LHC for $m_{Z^\prime}=3.5$ TeV and $g_{BL}=0.250$. 
The solid and dashed curves correspond to the results for $N(\nu_R)=2$ and $0$, respectively.  
The horizontal long-dashed line represents the SM cross section, 
  which is negligible compared with the $Z^\prime$ boson mediated process. 
}
\label{FigLHC}
\end{center}
\end{figure}

The differential cross section with respect to the final state dilepton invariant mass $M_{ll}$ is described as
\begin{eqnarray}
 \frac{d \sigma (pp \to \ell^+ \ell^- X) }
 {d M_{ll}}
 &=&  \sum_{a, b}
 \int^1_{-1} d \cos \theta
 \int^1_ \frac{M_{ll}^2}{E_{\rm CMS}^2} dx_1
 \frac{2 M_{ll}}{x_1 E_{\rm CMS}^2}   \nonumber \\
&\times & 
 f_a(x_1, Q^2)
  f_b \left( \frac{M_{ll}^2}{x_1 E_{\rm CMS}^2}, Q^2
 \right)  \frac{d \sigma(\bar{q} q \to \ell^+ \ell^-) }
 {d \cos \theta},
\label{CrossLHC}
\end{eqnarray}
where $E_{\rm CMS} =14$ TeV is the center-of-mass energy of the LHC.
In our numerical analysis, we employ CTEQ5M~\cite{Pumplin:2002vw} 
   for the parton distribution functions ($f_a$) with the factorization scale $Q= m_{Z^\prime}$. 
Reader may refer Appendix in Ref.~\cite{Iso:2009nw} for the helicity amplitudes 
   to calculate $d \sigma (\bar{q} q \to \ell^+ \ell^-)/d \cos \theta$.  
For the $Z^\prime$ boson mediated process, we consider two cases, 
   $N(\nu_R)=0$ and  $N(\nu_R)=2$, 
   where $N(\nu_R)$ is the number of right-handed (Dirac) neutrinos.  
For our case with $N(\nu_R)=2$, the total $Z^\prime$ boson decay width is given in Eq.~(\ref{width}), 
   while the number $15$ in the bracket must be replaced to $12$ for $N(\nu_R)=0$.

Fig.~\ref{FigLHC} shows the differential cross section for $pp \to e^+e^- X +\mu^+\mu^- X$ 
   for $m_{Z^\prime}=3.5$ TeV and $g_{BL}=0.250$,
   along with the SM cross section mediated by the $Z$-boson and photon (horizontal long-dashed line). 
The solid and dashed curves correspond to the results for $N(\nu_R)=2$ and $0$, respectively.  
The dependence of the total decay width on the number of right-handed neutrinos 
   reflects the resultant cross sections.  
When we choose a kinematical region for the invariant mass in the range,
   $M_{Z^\prime}- 100  \leq M_{ll}({\rm GeV}) \leq  M_{Z^\prime} + 100$, for example,   
   the signal events of $892$ and $1049$ for $N(\nu_R)=2$ and $0$, respectively, 
   would be observed with the prospective integrated luminosity of 1000/fb at the High-Luminosity LHC.   
The difference between $N(\nu_R)=2$ and $0$ are distinguishable with a $4-5\sigma$ significance.

\section{Conclusions and discussions} 

We have proposed a simple gauged U(1)$_{B-L}$ extension of the MSSM, 
   where R-parity is conserved as usual in the MSSM.  
The global $B-L$ symmetry in the MSSM is gauged and three right-handed neutrino chiral multiplets 
   are introduced, which make the model free from all gauge and gravitational anomalies. 
No $B-L$ Higgs field is introduced.   
We assign an even R-parity to one right-handed neutrino superfield $\Phi$, 
   while the other two right-handed neutrino superfields are odd as usual.  
The scalar component of $\Phi$ plays a role of the $B-L$ Higgs field to beak the U(1)$_{B-L}$ gauge symmetry   
   through its negative mass squared which is radiatively generated by the RG evolution 
   of soft SUSY breaking parameters. 
Therefore, the scale of the U(1)$_{B-L}$ symmetry breaking is controlled by the SUSY breaking parameters 
  and naturally be at the TeV scale.  
We have shown that this radiative symmetry breaking actually occurs with a suitable choice of  
   model parameters.   
Because of our novel R-parity assignment, three light neutrinos are Dirac particles with one massless state. 
Since R-parity is conserved, the lightest neutralino is a prime candidate of the dark matter of the universe. 
Depending on its mass, the lighter Majorana mass eigenstate ($\chi_\ell$) of a mixture of the $B-L$ gaugino and 
  the fermionic component of $\Phi$ (R-parity odd right-handed neutrino) appears as a new dark matter candidate.  
Assuming  $\chi_\ell$ is the lightest R-parity odd particle, we have calculated the dark matter relic abundance. 
When the mass of $\chi_\ell$ is close to half of the $Z^\prime$ boson mass,  
   the pair annihilation cross section of the dark matter particle is enhanced through the $Z^\prime$ boson resonance 
   in the $s$-channel process and the observed dark matter relic abundance is reproduced.   
We have also discussed LHC phenomenology for the Dirac neutrinos.   
The $Z^\prime$ boson, once discovered at the LHC,  will be a novel probe 
   of the Dirac nature of the light neutrinos since its invisible decay processes include 
   the final states with one massless (left-handed) neutrino and two Dirac neutrinos, 
   in sharp contrast with the conventional $B-L$ extension of the SM or MSSM, 
   where the right-handed neutrinos are heavy Majorana particles and decay 
   to the weak gauge bosons/Higgs boson plus leptons. 
If the $Z^\prime$ boson is discovered, the High-Luminosity LHC may reveal the existence of 
   the right-handed neutrino with a precise measurement of the total decay width of $Z^\prime$ boson.

Since the neutrinos are Dirac particles in our model, their Dirac Yukawa coupling must be extremely small. 
It is an important issue how to naturally realize such a small Yukawa coupling, 
  or a huge hierarchy between the neutrino Yukawa coupling and those of the other SM fermions,  
  in a reasonable theoretical framework. 
In addition, the mass squared hierarchy between $\phi$ and the other right-handed sneutrinos 
   is crucial to achieve the radiative $B-L$ gauge symmetry breaking. 
Realizing this hierarchy in a natural way is an additional issue.  
In order to solve these hierarchy problems, we may extend the model to the brane-world framework 
   with 5-dimensional warped space-time~\cite{Randall:1999ee}.   
Arranging the bulk mass parameters for the bulk hypermultiplets corresponding to matter and Higgs fields 
   in the minimal SUSY $B-L$ model, we can obtain large hierarchy among parameters 
   in 4-dimensional effective theory with mildly hierarchical  model parameters in the original 5-dimensional theory. 
This direction is worth investigating~\cite{OP2}.

\section*{Acknowledgments}
The work of N.O. is supported in part by the United States Department of Energy grant (DE-SC0013680).


\end{document}